\documentclass%
	[aps,pre,reprint,superscriptaddress,amsmath,amssymb,floatfix]%
	{revtex4-2}

\usepackage{amsopn}
\usepackage{enumerate,units,comment}
\usepackage{xcolor}
\usepackage{graphicx}
\usepackage[nooneline,raggedright,small,tight]{subfigure}
\usepackage{mathtools}
\usepackage{mathrsfs}

\allowdisplaybreaks[4]
\mathtoolsset{centercolon=true}

\figuretopcapfalse
\subfiguretopcaptrue
\DeclareMathOperator\diff{d\!}

\DeclareMathOperator\grad{grad}

\DeclareMathOperator\BigO{\mathcal{O}}

\def\cpi{\mathord{\mathrm{\pi}}}

\newcommand{\norm}[1]{\left\lVert{#1}\right\rVert}
\def\setsymb#1{\mathbb{#1}}

\def\ol#1{\overline{#1}}
\makeatletter
\newcommand{\subalign}[1]{%
  \vcenter{%
    \Let@ \restore@math@cr \default@tag
    \baselineskip\fontdimen10 \scriptfont\tw@
    \advance\baselineskip\fontdimen12 \scriptfont\tw@
    \lineskip\thr@@\fontdimen8 \scriptfont\thr@@
    \lineskiplimit\lineskip
    \ialign{\hfil$\m@th\scriptstyle##$&$\m@th\scriptstyle{}##$\crcr
      #1\crcr
    }%
  }
}
\makeatother
\newcommand\ignore[1]{}  

\begin{document}

\title{Phase description of chaotic oscillators}

\author{Takashi \surname{Imai}}
\email{takashi-imai@biwako.shiga-u.ac.jp}
\affiliation{The Center for Data Science Education and Research, Shiga University, Shiga 522-8522, Japan}
\author{Hiromichi \surname{Suetani}}
\affiliation{Faculty of Science and Technology, Oita University, Oita 870-1192, Japan}
\affiliation{Rhythm-Based Brain Information Processing Unit, RIKEN Center for Brain Science, Saitama 351-0198, Japan}
\author{Toshio \surname{Aoyagi}}
\affiliation{Graduate School of Informatics, Kyoto University, Kyoto 606-8501, Japan}

\date{\today}

\begin{abstract}
This paper presents a phase description of chaotic dynamics
for the study of chaotic phase synchronization.
A prominent feature of the proposed description
is that it systematically incorporates
the dynamics of the non-phase variables inherent in the system.
Taking these non-phase dynamics into account is essential
for capturing
the complicated nature of chaotic phase synchronization,
even in a qualitative manner.
We numerically verified the validity of the proposed description
in application to the R\"{o}ssler and Lorenz oscillators,
and we found that
our method provides an accurate description of
the characteristic distorted shapes of
the synchronization regions for these chaotic oscillators.
Furthermore,
the proposed description allows us to systematically identify and describe
the origin of this distortion.
\end{abstract}

\maketitle


\section{Introduction}\label{sec: Introduction}
The intrinsic rhythms exhibited by dynamical systems
have attracted interest in a wide range of fields
\cite{andronov1966theory,*gray1990chemical,*winfree2001geometry}.
For example,
the beating of the heart has been studied extensively,
not only because of its importance with regard to human health,
but also because it is a rich source of information as a dynamical system
\cite{glass1988clocks,*glass2001synchronization}.
In many cases,
such rhythmic systems interact with other oscillatory units,
and these interactions create further intriguing phenomena.
A typical example of such phenomena is phase synchronization
\cite{pikovsky2003synchronization},
for example,
synchronization between a heartbeat and locomotor rhythm
\cite{kirby1989coupling,*niizeki2005intramuscular},
which is thought to
improve the efficiency of blood circulation through active muscles.

The phase reduction approach provides a systematic method for analyzing phase synchronization
\cite{winfree1967biological,*kuramoto1984chemical,*hoppensteadt1997weakly,*strogatz2000fromkuramoto,*izhikevich2007dynamical}.
This method provides
a concise description of rhythm dynamics,
and it has served as a framework for the study of phase synchronization
for many years.
In this way, the phase reduction approach has contributed greatly
to our understanding of phase synchronization phenomena.

Although, in its conventional form, the phase reduction approach can be
applied only to weakly perturbed limit-cycle oscillators,
recently this approach has been extended to a more general form with broader application.
For example,
Refs.~\cite{yoshimura2008phase,*teramae2009stochastic,*goldobin2010dynamics}
and \cite{nakao2012phase,*nakao2014phase}
demonstrate that
phase reduction can be extended to
noisy limit-cycle oscillators
and
limit-cycle solutions of reaction-diffusion systems,
respectively.
However, the application of phase reduction to the analysis of chaotic oscillators%
---a very common type of rhythmic system---%
has not yet been established.

For chaotic oscillators,
the emergence of a variant of phase synchronization can often be found
when the behavior of the system is described
in terms of properly defined phase variables
\cite{stone1992frequency,*rosenblum1996phase}.
This type of synchronization phenomenon exhibited by
the phase variables in descriptions of chaotic dynamics
is called \emph{chaotic phase synchronization} (CPS).
We believe that
establishing the application of the phase reduction approach to the analysis of chaotic oscillators
would lead to significant progress in our understanding of CPS.
In this paper, we present formalism that does indeed accomplish this.

A particularly difficult problem
in formulating the phase reduction analysis of chaotic oscillators
is to incorporate a proper treatment of the non-phase variables.
For chaotic systems,
in general,
even a weak perturbation can cause
a qualitative change in the behavior of the non-phase variables.
Such changes may drastically alter the rhythmic properties of the oscillator.
For this reason,
it is important to
properly treat the dynamics of the non-phase variables.
In this regard,
there is considerable room for improvement
in the approaches
proposed in previous studies on the phase reduction of chaotic oscillators
\cite{pikovsky1997phasephysicad,pikovsky1997phasechaos,josic2001phase,beck2003geometric,tonjes2017phase}.
For example,
the phase description proposed in Ref.~\cite{pikovsky1997phasechaos}
does not include the perturbation dependence of the non-phase variables,
and that proposed in Ref.~\cite{pikovsky1997phasephysicad}
does not decouple the non-phase dynamics from the phase dynamics.
Contrastingly, in this paper
we construct a phase description of chaotic dynamics
that systematically incorporates
the dynamics of the non-phase variables.

A key step toward constructing the phase description is
to define a phase variable
with which the system dynamics can be expressed in the desired form.
Some studies on CPS define the phase variable as a simple geometric angle of the state vector,
such as the azimuthal angle in the three-dimensional state space.
While this type of phase variable has the advantage
that its properties are relatively well understood
(see, e.g., Refs.~\cite{pereira2007phase,*pereira2008phase}),
it is unsuitable for the phase description, because its use results in
phase and non-phase dynamics that are too closely coupled.
More suitable phase variables are introduced in
Refs.~\cite{josic2001phase,schwabedal2012optimal,tonjes2017phase},
but these are still inadequate for our purposes,
because with them,
the dynamics of the non-phase variables cannot be expressed
in sufficiently explicit forms to allow examination of
the influence of a perturbation on the behavior of the non-phase variables.
In this paper, we propose yet another definition of a phase variable.

\section{Phase Description of Chaotic Dynamics}\label{sec: Phase Description of Chaotic Dynamics}
\begin{figure}[tp]
\includegraphics{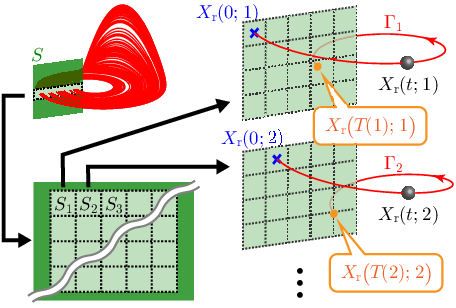}
\caption{\label{fig: representative trajectory}%
(Color online)
Construction of $\left\{S_i\right\}$ and $\left\{\Gamma_i\right\}$.
}
\end{figure}

To derive a phase description of chaotic dynamics,
first, we consider an unperturbed system of the form
	\begin{equation}\label{eq: free oscillator}
	\dot{X}
	=	F(X),\quad
	X
	\in	\setsymb{R}^{m},
	\end{equation}
which we assume to possess a phase-coherent chaotic attractor
\cite{beck2003geometric}, $A$.
Let $S$ be an $\left(m - 1\right)$-dimensional surface of section
transverse to $A$
such that all trajectories starting on $S$ return to $S$
within a certain period of time.
This surface can be partitioned into
(nonempty, pairwise disjoint, covering) small cells,
$S_i$.
For each cell $S_i$,
we choose a solution $X_\text{r}(t;\,i)$
of the differential equation $\eqref{eq: free oscillator}$,
with the initial condition $X_\text{r}(0;\,i) \in S_i$.
The solution $X_\text{r}(t;\,i)$
passes through the surface of section $S$ repeatedly
after the initial time ($t = 0$),
and thus the time $T(i)$ of its first return to $S$ can be defined
(see Fig.~\ref{fig: representative trajectory}).
We call each trajectory $
	\Gamma_i
	:=	\left\{
			X_\text{r}(t;\,i)
			\mid
			0 \le t \le T(i)
		\right\}
$ the \emph{representative trajectory} for the cell $S_i$,
and we employ a set of the representative trajectories,
$\left\{\Gamma_i\right\}$,
as reference orbits for introducing a phase variable.

Suppose that in an open neighborhood $U$ of the attractor $A$,
there exists a change of coordinates
	\[
	\begin{array}{ccc}
	U
	&	\xrightarrow{\hspace{2em}}
		&	\setsymb{R}^{m-1} \times \left[0,\,2 \cpi\right]\\[.5ex]
	\rotatebox{90}{$\in$}
		&
			&	\rotatebox{90}{$\in$}\\[.5ex]
	X
	&	\xmapsto{\hspace{2em}}
		&	\bigl(R(X),\,\phi(X)\bigr)
	\end{array}
	\]
such that
$R$ is smooth in $U$,
$\phi$ is continuous in $U$ and smooth in $U \setminus S$,
$\phi(X) = 0$ if and only if $X \in S \cap U$,
and
	\begin{equation}
	{\diff \phi\bigl(X_\text{r}(t;\,i)\bigr) \over \diff t}
	=	\omega(i)
	:=	{2 \cpi \over T(i)}
	\quad\text{if\ \ $0 < t < T(i)$},
	\label{eq: condition of uniform evolution}
	\end{equation}
where the endpoints of the interval $\left[0,\,2 \cpi\right]$
are identified with each other.
Hereafter,
we write $R\bigl(X(t)\bigr)$ and $\phi\bigl(X(t)\bigr)$
simply as $R(t)$ and $\phi(t)$.
Although,
in terms of the coordinates $\left(R,\,\phi\right)$,
the system $\eqref{eq: free oscillator}$ can be formally expressed as
	\begin{equation}
	\dot{R}(t)
	=	F_R\bigl(R(t),\,\phi(t)\bigr),\quad
	\dot{\phi}(t)
	=	F_\phi\bigl(R(t),\,\phi(t)\bigr),
	\label{eq: free evolution of phase - prototype}
	\end{equation}
it is more convenient to rewrite the latter in a phase-oscillator-like form.
Let $\mathscr{T}_n$ be the time of the $n$th return to $S$,
and
let $\mathscr{C}_n$ be the index of the cell
in which the state of the system exists at $\mathscr{T}_n$.
For $t \in \left(\mathscr{T}_n,\,\mathscr{T}_{n+1}\right)$,
the time evolution of $\phi$ can be expressed as follows:
	\begin{equation}\label{eq: free evolution of phase}
	\dot{\phi}(t)
	=	\omega(\mathscr{C}_n)
		+
		\rho\bigl(R(t),\,\phi(t)\bigr).
	\end{equation}
When $t$ passes $\mathscr{T}_{n+1}$
(in other words, the next time the state returns to $S$),
the evolution equation for $\phi$ is replaced with
the equation corresponding to the next cell, $\mathscr{C}_{n+1}$.
The condition $\eqref{eq: condition of uniform evolution}$
implies the condition
	\begin{equation}\label{eq: rho on representative trajectory}
	\rho\bigl(R_\text{r}(t - \mathscr{T}_n;\,\mathscr{C}_n),\,\phi(t)\bigr)
	=	0,
	\end{equation}
where $
	R_\text{r}(t;\,i)
	:=	R\bigl(X_\text{r}(t;\,i)\bigr)
$.

Next, we consider the situation in which
this oscillator is subject to a weak perturbation, $\varepsilon p(t)$,
and the system is governed by the equation
	\[
	\dot{X}(t)
	=	F\bigl(X(t)\bigr) + \varepsilon p(t).
	\]
In this situation,
the extent to which the time evolutions of $R$ and $\phi$
are perturbed
depends on sensitivity functions.
Explicitly, these time evolutions are described by the following:
	\begin{align}
	\dot{R}(t)
	=	{}
	&	F_R\bigl(R(t),\,\phi(t)\bigr)
		+
		\varepsilon
		\left.{\partial R(X) \over \partial X}\right\rvert_{X = X(t)}
			p(t),
	\label{eq: perturbed evolution of amplitude}\\
	\dot{\phi}(t)
	=	{}
	&	\omega(\mathscr{C}_n)
		+
		\rho\bigl(R(t),\,\phi(t)\bigr)
	\notag\\*
	&	{} +
		\varepsilon \left.\grad_{X} \phi(X) \right\rvert_{X = X(t)}
			\cdot p(t).
	\label{eq: perturbed evolution of phase}
	\end{align}

Let $\delta R(t)$ represent the deviation of $R(t)$
from its value on the representative trajectory;
i.e., $
	\delta R(t)
	:=	R(t) - R_\text{r}(t - \mathscr{T}_n;\,\mathscr{C}_n)
$ for $t \in \left(\mathscr{T}_n,\,\mathscr{T}_{n+1}\right]$.
Here, we assume that
the surface of section, $S$, is partitioned
so finely that
$\norm{\delta R(t)}$ can be regarded as a small quantity
of $\BigO(\varepsilon)$
until the next return to $S$.
Ignoring
terms of second and higher order in $\varepsilon$
and using the condition $\eqref{eq: rho on representative trajectory}$,
we can rewrite
Eqs.~$\eqref{eq: perturbed evolution of amplitude}$ and $\eqref{eq: perturbed evolution of phase}$
as
	\begin{align}
	\delta \dot{R}(t)
	=	{}
	&	\left.
			{\partial F_R \over \partial R}
		\right\rvert_{
			\subalign{
				R
				&=	R_\text{r}(t - \mathscr{T}_n;\,\mathscr{C}_n)\\
				\phi
				&=	\omega(\mathscr{C}_n) \left(t - \mathscr{T}_n\right)
			}
		}
			\delta R(t)
	\notag\\*
	&{}	+
		\varepsilon \left.
			{\partial R \over \partial X}
		\right\rvert_{
			X
			=	X_\text{r}(t - \mathscr{T}_n;\,\mathscr{C}_n)
		}
			p(t),
	\label{eq: neglect small terms of perturbed system - amplitude}\\
	\dot{\phi}(t)
	=	{}
	&	\omega\bigl(\mathscr{C}_n\bigr)
		+
		\left.\grad_{R} \rho(R,\,\phi) \right\rvert_{
			\subalign{
				R
				&=	R_\text{r}(t - \mathscr{T}_n;\,\mathscr{C}_n)\\
				\phi
				&=	\omega(\mathscr{C}_n) \left(t - \mathscr{T}_n\right)
			}
		}
			\cdot \delta R(t)
	\notag\\*
	&{}	+
		\varepsilon \zeta\bigl(\omega(\mathscr{C}_n) \left(t - \mathscr{T}_n\right);\,\mathscr{C}_n\bigr) \cdot p(t),
	\label{eq: neglect small terms of perturbed system - phase}
	\end{align}
where
	\begin{align*}
	\zeta(\phi;\,i)
	&:=	\left.\grad_{X} \phi(X) \right\rvert_{
			X
			=	X_\text{r}[\phi / \omega(i);\,i]
		}.
	\end{align*}

Define $
	\delta R_n
	:=	\lim_{t \downarrow \mathscr{T}_n} \delta R(t)
	=	R_n - R_\text{r}(0;\,\mathscr{C}_n)
$,
where $R_n := R(\mathscr{T}_n)$.
Then for $t \in \left(\mathscr{T}_n,\,\mathscr{T}_{n+1}\right]$,
the differential equation $\eqref{eq: neglect small terms of perturbed system - amplitude}$
has the solution
	\begin{multline}
	\delta R(t)
	=	\Phi(t - \mathscr{T}_n;\,\mathscr{C}_n)
		\left[
			\delta R_n
			\vphantom{
			+
			\varepsilon \int_{\mathscr{T}_n}^{t}
				\Upsilon(s - \mathscr{T}_n;\,\mathscr{C}_n)
				p(s)
				\diff s
			}
		\right.\\*
		\left.
			\vphantom{
			\delta R_n
			}
			+
			\varepsilon \int_{\mathscr{T}_n}^{t}
				\Upsilon(s - \mathscr{T}_n;\,\mathscr{C}_n)
				p(s)
				\diff s
		\right],
	\label{eq: R-deviation from representative trajectory}
	\end{multline}
where $\Phi(t;\,i)$ is the fundamental matrix of the homogeneous equation
	\[
	\delta \dot{R}(t)
	=	\left.
			{\partial F_R \over \partial R}
		\right\rvert_{
			\subalign{
				R
				&=	R_\text{r}(t;\,i)\\
				\phi
				&=	\omega(i) t
			}
		}
			\delta R(t)
	\]
whose value at the initial time, $\Phi(0;\,i)$, is the identity matrix,
and
	\[
	\Upsilon(t;\,i)
	:=	\Phi^{-1}(t;\,i)
		\left.
			{\partial R \over \partial X}
		\right\rvert_{
			X
			=	X_\text{r}(t;\,i)
		}.
	\]
From Eq.~$\eqref{eq: R-deviation from representative trajectory}$,
we obtain the recurrence equation
	\begin{alignat}{2}
	R_{n+1}
	&{} \simeq {}
	&	R_\text{r}\bigl(T(\mathscr{C}_n);\,\mathscr{C}_n\bigr)
		+
		\Phi\bigl(T(\mathscr{C}_n);\,\mathscr{C}_n\bigr)
		\left[
			\delta R_n
			\vphantom{
			+
			\varepsilon \int_{\mathscr{T}_n}^{\mathscr{T}_{n+1}}
				\Upsilon(t - \mathscr{T}_n;\,\mathscr{C}_n)
				p(t)
				\diff t
			}
		\right.\quad
	&
	\notag\\*
	&
	&	\left.
			\vphantom{
			\delta R_n
			}
			+
			\varepsilon \int_{\mathscr{T}_n}^{\mathscr{T}_{n+1}}
				\Upsilon(t - \mathscr{T}_n;\,\mathscr{C}_n)
				p(t)
				\diff t
		\right]
	&	{}.
	\label{eq: map dynamics}
	\end{alignat}

The second term on the right-hand side of
Eq.~$\eqref{eq: neglect small terms of perturbed system - phase}$
represents the deviation of the frequency
from its value on the representative trajectory,
which is caused by $\delta R(t)$.
Now, assume that the surface of section is selected
to be one for which the variation in the return time is very small.
Specifically, we assume that
this variation is sufficiently small that
the second term on the right-hand side
of Eq.~$\eqref{eq: neglect small terms of perturbed system - phase}$
is much smaller than the third term.
Ignoring this small term,
Eq.~$\eqref{eq: neglect small terms of perturbed system - phase}$ can be rewritten as
	\begin{equation}\label{eq: reduced phase equation (neglecting nonuniformity of frequency)}
	\dot{\phi}(t)
	=	\omega(\mathscr{C}_n)
		+
		\varepsilon \zeta\bigl(\phi(t);\,\mathscr{C}_n\bigr) \cdot p(t).
	\end{equation}

Equation $\eqref{eq: reduced phase equation (neglecting nonuniformity of frequency)}$
has the same form as the phase oscillator model,
except for the $\mathscr{C}_n$-dependence.
To remove the $\mathscr{C}_n$-dependence from Eq.~$\eqref{eq: reduced phase equation (neglecting nonuniformity of frequency)}$,
a bit more consideration is necessary.
For simplicity, let us restrict the class of perturbations
to periodic driving functions.
With a weak periodic driving function $\varepsilon \tilde{p}(\theta)$
whose angular frequency, $\Omega$, is close to
the average frequency of the unperturbed system $\eqref{eq: free oscillator}$,
the time evolution of the phase difference, $
	\psi(t)
	:=	\phi(t) - \Omega t
$, will be much slower than that of $R(t)$.
In this case, it is reasonable to regard $\psi$ as a constant
on the time scale of the $R$-dynamics.
This allows us to reduce Eq.~$\eqref{eq: map dynamics}$ to the map
	\begin{align}
	R_{n+1}
	\simeq
		{}
	&	R_\text{r}\bigl(T(\mathscr{C}_n);\,\mathscr{C}_n\bigr)
	\notag\\*
	&{}	+
		\Phi\bigl(T(\mathscr{C}_n);\,\mathscr{C}_n\bigr)
			\left[
				\delta R_n
				+
				\varepsilon H(\mathscr{C}_n,\,\psi)
			\right],
	\label{eq: reduced map dynamics}
	\end{align}
where
$H(i,\,\psi)$ denotes the averaged effect of the periodic driving:
	\[
	H(i,\,\psi)
	=	{1 \over \omega(i)}
		\int_{0}^{2 \cpi}
			\Upsilon\mathopen{}\left[{\phi \over \omega(i)};\,i\right]\mathclose{}
			\tilde{p}(\phi - \psi)
			\diff \phi.
	\]
The chaotic behavior of the system is encapsulated
in the map $\eqref{eq: reduced map dynamics}$.
Iteration of Eq.~$\eqref{eq: reduced map dynamics}$ produces
the (conditional) natural measure $\mu_i(\psi;\,\varepsilon)$
\cite{farmer1983dimension,*blank2003multicomponent},
the probability of visiting the cell $S_i$
under the condition that the phase difference is equal to $\psi$.
Here, we ignore the small fluctuations in the actual phase difference $\psi(t)$
produced by the slight differences in $\omega(i)$
among the representative trajectories,
and focus on the drift component of $\psi(t)$.
Inside and near the synchronization region, the time scale characterizing the evolution of this drift component will be much longer
than that characterizing the convergence of the relative frequency distribution of $S_i$
to the natural measure, $\mu_i$.
Averaging the $\psi$-dynamics with respect to $\mu_i$,
we obtain
from Eq.~$\eqref{eq: reduced phase equation (neglecting nonuniformity of frequency)}$
flow described by
	\begin{equation}\label{eq: flow dynamics}
	\dot{\psi}(t)
	=	\delta \omega\bigl(\psi(t);\,\varepsilon\bigr)
		+
		\varepsilon G\bigl(\psi(t);\,\varepsilon\bigr),
	\end{equation}
where
	\begin{equation}\label{eq: adjuster}
	\left.
	\begin{gathered}
	\delta \omega(\psi;\,\varepsilon)
	:=	\ol{\omega}(\psi;\,\varepsilon) - \Omega,\quad
	\ol{\omega}(\psi;\,\varepsilon)
	:=	{2 \cpi \over \ol{T}(\psi;\,\varepsilon)},\\
	\ol{T}(\psi;\,\varepsilon)
	:=	\sum_{i} \mu_i(\psi;\,\varepsilon) T(i),\\
	G(\psi;\,\varepsilon)
	:=	{1 \over 2 \cpi}
		\int_{0}^{2 \cpi}
			\ol{\zeta}(\phi,\,\psi;\,\varepsilon) \cdot \tilde{p}(\phi - \psi)
			\diff \phi,\\
	\ol{\zeta}(\phi,\,\psi;\,\varepsilon)
	:=	\sum_{i}
			\mu_i(\psi;\,\varepsilon)
			{T(i) \over \ol{T}(\psi;\,\varepsilon)}
			\zeta(\phi;\,i).
	\end{gathered}
	\,\right\}
	\end{equation}
This averaging may not be valid in a rigorous sense,
because of the singularity of $\mu_i$ as a function of $\psi$.
However, it can provide a rough approximation of the original system,
as demonstrated in the following section.
Note that the flow described by Eq.~$\eqref{eq: flow dynamics}$ does not involve $\mathscr{C}_n$.

Several phase equations of the same form as Eq.~$\eqref{eq: reduced phase equation (neglecting nonuniformity of frequency)}$
have been presented in previous studies
(e.g., Ref.~\cite{pikovsky1997phasephysicad}).
Here we have derived a reduced version of this description
by separating the fast and slow dynamics
(i.e., the dynamics of $R$ and $\psi$, respectively).
This new description
allows us to analyze the behavior of $R$ separately from the phase dynamics.
[Note that $\psi$ in Eq.~$\eqref{eq: reduced map dynamics}$ is a constant parameter
whose value can be set independently of
the state $\psi(t)$ of the flow $\eqref{eq: flow dynamics}$.]
For this reason, it should be helpful in elucidating
how the phase dynamics are affected
by a change in the chaotic behavior of the non-phase variables.

\begin{figure}[tp]
\includegraphics{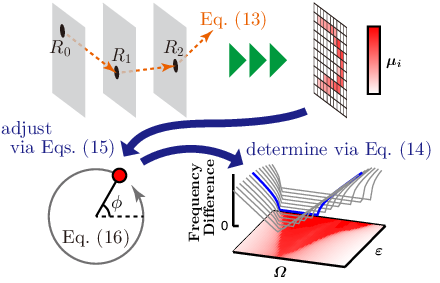}
\caption{\label{fig: MAPO model}%
(Color online)
Representation using the map-adjusted phase oscillator (MAPO) model
depicting how the rhythmic properties of the chaotic oscillator are determined.
}
\end{figure}

What does the existence of the flow
described by Eq.~$\eqref{eq: flow dynamics}$ indicate?
Note that Eq.~$\eqref{eq: flow dynamics}$ can be viewed
as an averaged equation derived from the phase oscillator
	\begin{equation}\label{eq: underlying phase oscillator}
	\dot{\phi}(t)
	=	\ol{\omega}(\psi;\,\varepsilon)
		+
		\varepsilon \ol{\zeta}\bigl(\phi(t),\,\psi;\,\varepsilon\bigr) \cdot \tilde{p}(\Omega t),
	\end{equation}
which is adjusted by the map $\eqref{eq: reduced map dynamics}$
in the sense that
$\ol{\omega}$ and $\ol{\zeta}$ depend on $\mu_i$
via Eqs.~$\eqref{eq: adjuster}$.
In other words,
the rhythm dynamics of chaotic oscillators
are described as a \emph{map-adjusted phase oscillator} (MAPO).
The existence of the flow described by Eq.~$\eqref{eq: flow dynamics}$
therefore indicates that
the MAPO $\eqref{eq: underlying phase oscillator}$
with the map $\eqref{eq: reduced map dynamics}$ determines
the rhythmic properties (such as the average frequency)
of the original chaotic oscillator,
as illustrated in Fig.~\ref{fig: MAPO model}.

\section{Numerical Examples}\label{sec: Numerical Examples}
In this section,
we demonstrate the validity of the description proposed here
through consideration of numerical examples.

\subsection{The R\"{o}ssler oscillator}
\begin{figure*}[tp]
\includegraphics{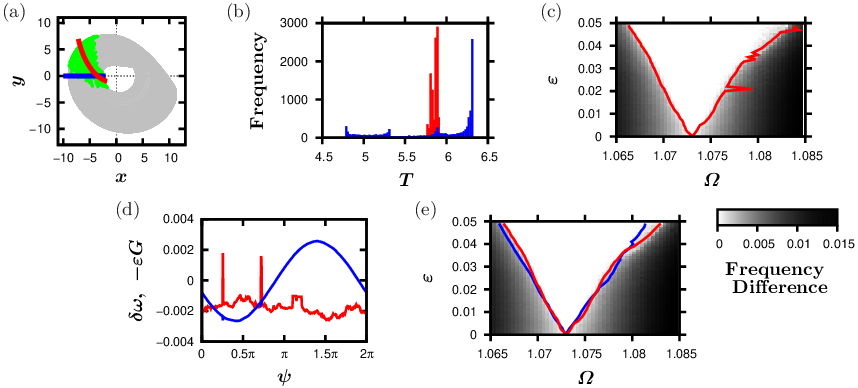}
\caption{\label{fig: demonstration on Rossler oscillator}%
(Color online)
Numerical results for the R\"{o}ssler oscillator.
\subref{fig: demonstration on Rossler oscillator - optimal isophase}
The trajectory (gray curve),
the stroboscopic set \protect\cite{schwabedal2012optimal} (green dots),
and the adopted surface of section (red curve).
The blue line segment represents
a simpler
surface of section ($x < 0$, $y = 0$).
These surfaces of section yield the return time distributions
plotted in \subref{fig: demonstration on Rossler oscillator - return time distribution},
where
the red and blue histograms depict the distributions
for the former and latter surfaces of section, respectively.
\subref{fig: demonstration on Rossler oscillator - tongue}
The difference
between the frequencies of the periodic driving and the driven system
observed in the full model.
The red curve indicates
the edge of the Arnold tongue derived from the MAPO model
[i.e., the boundary of
the $\left(\varepsilon,\,\Omega\right)$-region in which CPS occurs
in the MAPO model].
\subref{fig: demonstration on Rossler oscillator - characteristic curves}
The instantaneous frequency difference $\delta \omega$ (red curve)
and the coupling function $G$ (blue curve)
at $\left(\varepsilon,\,\Omega\right) = \left(0.021,\,1.077\right)$.
By preprocessing these quantities using a moving average filter,
we obtained
a more accurate approximation
\subref{fig: demonstration on Rossler oscillator - tongue without prickles}.
In this figure, for reference,
the blue curve indicates the edge of the Arnold tongue
obtained by using the other surface of section.
}
\makeatletter
\let\subfig@oldlabel=\label
\let\label=\subfloat@label
\makeatother
\refstepcounter{subfigure}\label{fig: demonstration on Rossler oscillator - optimal isophase}
\refstepcounter{subfigure}\label{fig: demonstration on Rossler oscillator - return time distribution}
\refstepcounter{subfigure}\label{fig: demonstration on Rossler oscillator - tongue}
\refstepcounter{subfigure}\label{fig: demonstration on Rossler oscillator - characteristic curves}
\refstepcounter{subfigure}\label{fig: demonstration on Rossler oscillator - tongue without prickles}
\end{figure*}

As a first example,
we consider the following R\"{o}ssler oscillator \cite{rossler1976equation}
driven by a weak sinusoidal perturbation:
	\begin{equation}\label{eq: forced Rossler equation}
	\begin{gathered}
	\dot{x}
	=	-y - z + \varepsilon \sin \Omega t,\\
	\dot{y}
	=	x + a y,\quad
	\dot{z}
	=	b + z \left(x - c\right),
	\end{gathered}
	\end{equation}
with the parameter values $a = 0.2$, $b = 0.2$, and $c = 5.7$.
Does the MAPO model consisting of
Eqs.~$\eqref{eq: reduced map dynamics}$ and $\eqref{eq: underlying phase oscillator}$
accurately produce the Arnold tongue for this system?
The Arnold tongue can be constructed using the MAPO model
by calculating
the range of $\Omega$
in which there exists at least one value of $\psi$
satisfying the conditions
	\begin{equation}\label{eq: conditions of stable fixed point}
	f_\varepsilon(\psi)
	=	0\quad
	\text{and}\quad
	{\diff f_\varepsilon \over \diff \psi}(\psi)
	<	0,
	\end{equation}
where
$f_\varepsilon(\psi)$ denotes
the right-hand side of Eq.~$\eqref{eq: flow dynamics}$%
---i.e.,
	\[
	f_\varepsilon(\psi)
	=	\delta \omega(\psi;\,\varepsilon)
		+
		\varepsilon G(\psi;\,\varepsilon).
	\]
(For chaotic systems,
averaged quantities, such as $\delta \omega$ and $G$,
can be non-differentiable
\cite{ershov1993perturbation,*keller2008continuity,*baladi2014linear}.
For this reason, in general, more careful consideration is needed.)
To obtain $\delta \omega$ and $G$,
we adopt the surface of section
depicted in Fig.~\ref{fig: demonstration on Rossler oscillator - optimal isophase},
which is one of the optimal isophases
(constructed using the method introduced in Ref.~\cite{schwabedal2012optimal})
of the unperturbed system.
With this surface of section,
the variation in the return time
[see Fig.~\ref{fig: demonstration on Rossler oscillator - return time distribution}]
is sufficiently small that
Eq.~$\eqref{eq: reduced phase equation (neglecting nonuniformity of frequency)}$
provides an accurate approximation of
Eq.~$\eqref{eq: neglect small terms of perturbed system - phase}$.
Using numerical simulations,
we can easily evaluate $T(i)$, $R_\text{r}\bigl(T(i);\,i\bigr)$,
$\Phi\bigl(T(i);\,i\bigr)$, $\Upsilon(t;\,i)$,
and $\zeta(\phi;\,i)$.
This allows us to calculate
$\delta \omega$ and $G$ defined in Eqs.~$\eqref{eq: adjuster}$.
With this treatment,
the MAPO model produces the Arnold tongue
depicted in Fig.~\ref{fig: demonstration on Rossler oscillator - tongue}.
Except in several isolated regions,
the discrepancy between the form of
the Arnold tongue derived from the MAPO model
and that derived from the original model
is small.
The only significant discrepancy between the two
consists of several horizontal spikes in the former
[e.g., near $\left(\varepsilon,\,\Omega\right) = \left(0.021,\,1.077\right)$].

To understand the appearance of the horizontal spikes,
let us consider Fig.~\ref{fig: demonstration on Rossler oscillator - characteristic curves},
which plots
the instantaneous frequency difference, $\delta \omega$,
and the coupling function, $G$, on one of these spikes.
The curves in this figure have jumps at several values of $\psi$.
These jumps greatly extend the range of $\Omega$
in which there exists at least one value of $\psi$
satisfying the conditions $\eqref{eq: conditions of stable fixed point}$.
This leads to the emergence of the spike.
Closer investigation reveals that
these jumps correspond to the values of $\psi$ at which
narrow periodic windows appear
in the $R$-dynamics $\eqref{eq: reduced map dynamics}$.
The spikes reflect
these large changes occurring in narrow parameter ranges.

The spikes are not observed in the Arnold tongue derived from the full model.
This is because, in the full model,
the phase difference $\psi$ fluctuates slightly
on the time scale of the $R$-dynamics,
and consequently $R$ does not remain confined to these narrow windows.
This observation suggests that
the fast fluctuations in $\psi$ play a key role
in smoothing the frequency change.

To produce an effect similar to that caused by the fast fluctuations in $\psi$,
we can utilize a moving average filter.
By applying this filter to $\delta \omega$ and $G$,
we can adjust the MAPO model
so that it provides a more accurate approximation
[see Fig.~\ref{fig: demonstration on Rossler oscillator - tongue without prickles}].

The remaining slight difference between
the Arnold tongues obtained from the full model
and the MAPO model
can be attributed to
the anomalous enhancement of the diffusion coefficient
discussed in Ref.~\cite{fujisaka2005chaotic}.
This enhancement occurs near the point
at which CPS breaks down
and weakens the system's periodicity in that region.
Because the MAPO model is constructed
assuming strong periodicity of the system,
as this periodicity weakens,
the discrepancy between
the forms of the Arnold tongues derived from the full model and the MAPO model
increases.
However, as long as it is not too large,
the error inherent in the form derived from the MAPO model
does not prevent us from obtaining an accurate understanding
of the global structure of the tongue,
because this error is localized
in the region near the point at which CPS breaks down.

\subsection{The Lorenz oscillator}
\begin{figure*}[tp]
\includegraphics{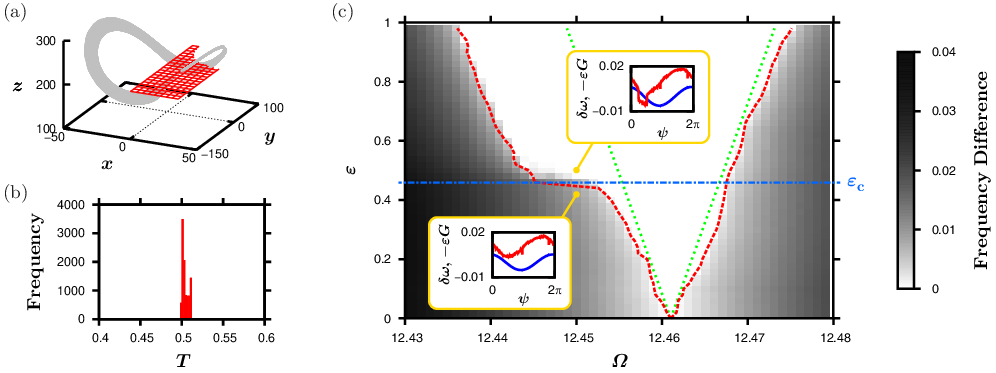}
\caption{\label{fig: demonstration on Lorenz oscillator}%
(Color online)
Numerical results for the Lorenz oscillator.
\subref{fig: demonstration on Lorenz oscillator - trajectory and isophase}
The trajectory (gray curve)
and the adopted surface of section (red mesh surface).
This surface of section yields the return time distribution
plotted in \subref{fig: demonstration on Lorenz oscillator - return time distribution}.
\subref{fig: demonstration on Lorenz oscillator - tongue}
The difference
between the frequencies of the periodic driving and the driven system
observed in the full model.
The dashed red curve indicates
the edge of the Arnold tongue obtained from the MAPO model
using the smoothed $\delta \omega$ and $G$.
For reference,
the edge of the tongue obtained
by ignoring the perturbation dependence of the non-phase variables
[i.e., by using $\delta \omega(\psi;\,0)$ and $G(\psi;\,0)$
instead of $\delta \omega(\psi;\,\varepsilon)$ and $G(\psi;\,\varepsilon)$]
is indicated by the dotted green curve.
The insets present
representative examples of $\delta \omega$ (red curve) and $-\varepsilon G$ (blue curve)
calculated in the regions above and below $\varepsilon = \varepsilon_\text{c}$,
where $\varepsilon_\text{c} \approx 0.46$ is the value of $\varepsilon$
at which the Arnold tongue for the Lorenz oscillator exhibits an abrupt extension
along the $\Omega$-axis.
The crossing of $\delta \omega$ and $-\varepsilon G$ indicates that CPS occurs
at the values of $\varepsilon$ and $\Omega$ used to plot their curves.
}
\makeatletter
\let\subfig@oldlabel=\label
\let\label=\subfloat@label
\makeatother
\refstepcounter{subfigure}\label{fig: demonstration on Lorenz oscillator - trajectory and isophase}
\refstepcounter{subfigure}\label{fig: demonstration on Lorenz oscillator - return time distribution}
\refstepcounter{subfigure}\label{fig: demonstration on Lorenz oscillator - tongue}
\end{figure*}

As a second example,
we consider the following Lorenz oscillator \cite{lorenz1963deterministic}
driven by a weak sinusoidal perturbation:
	\begin{equation}\label{eq: forced Lorenz equation}
	\begin{gathered}
	\dot{x}
	=	\sigma \left(y - x\right) + \varepsilon \sin \Omega t,\\
	\dot{y}
	=	x \left(\rho - z\right) - y,\quad
	\dot{z}
	=	x y - \beta z,
	\end{gathered}
	\end{equation}
with the parameter values $\sigma = 10$, $\beta = 8/3$, and $\rho = 210$
\cite{park1999phase}.
We use the flat surface depicted in Fig.~\ref{fig: demonstration on Lorenz oscillator - trajectory and isophase}
as the surface of section.
Although this surface was not determined through a careful optimization procedure,
with it, the variation of the return time is indeed small,
varying by only approximately $\pm 2 \%$
[see Fig.~\ref{fig: demonstration on Lorenz oscillator - return time distribution}].
(This is of similar magnitude to the variation
observed in our investigation of the R\"{o}ssler oscillator, discussed above,
in which we used an optimal isophase.)
With this small variation,
the form of the Arnold tongue derived from the MAPO model in this case
again deviates only slightly from that derived from the full model,
as shown in Fig.~\ref{fig: demonstration on Lorenz oscillator - tongue}.

The Arnold tongue for the Lorenz oscillator $\eqref{eq: forced Lorenz equation}$
exhibits an abrupt extension
along the $\Omega$-axis
at $\varepsilon = \varepsilon_\text{c} \approx 0.46$.
This extension,
like the spikes in the case of the R\"{o}ssler oscillator,
also reflects the appearance of periodic windows in the $R$-dynamics.
The difference between the present situation and that for the R\"{o}ssler oscillator
is that in the present situation, the periodic windows are so wide that
despite the fluctuations of $\psi$,
$R$ remains confined to these windows for some time,
and hence these windows have a strong effect
even on the behavior of the full model.
(Indeed,
the Lorenz oscillator exhibits periodic behavior
in part of the region above $\varepsilon = \varepsilon_\text{c}$,
as reported in Ref.~\cite{park1999phase}.)
Note that
the MAPO model accurately approximates the frequency
in the regions both above and below $\varepsilon = \varepsilon_\text{c}$.
This indicates that
the MAPO model properly inherits
the perturbation dependence of the non-phase variables
from the full model.
This inheritance results from the fact that
$\ol{\omega}$ and $\ol{\zeta}$
in the phase oscillator $\eqref{eq: underlying phase oscillator}$
depend on the natural measure $\mu_i$;
when the periodic driving triggers
a qualitative change in the behavior of $R$,
this change leads to large changes in $\mu_i$,
and thus in the characteristics of
the phase oscillator $\eqref{eq: underlying phase oscillator}$
[see the insets of Fig.~\ref{fig: demonstration on Lorenz oscillator - tongue}].
In this way,
the MAPO model successfully incorporates
the dynamics of the non-phase variables existing in the system.

\section{Conclusion}\label{sec: Conclusion}
In summary,
we have proposed the MAPO model as a framework for the study of CPS.
This model consists of
a chaotic map describing the evolution of the non-phase variables
and a phase oscillator adjusted
in accordance with the natural measure of the map,
as illustrated in Fig.~\ref{fig: MAPO model}.
This map, as well as the phase oscillator,
depends on the perturbation
applied to the original chaotic oscillator.
Accordingly,
the MAPO model allows us to examine
how the perturbation induces
a qualitative change in the behavior of the non-phase variables
and alters the rhythmic properties of the chaotic oscillator.

In general, the Arnold tongues of chaotic oscillators
may have distorted shapes different from
the triangular shape exhibited by the tongues of limit-cycle oscillators.
The origin of this distortion
is clarified through analysis of the MAPO model,
as demonstrated in Sec.~\ref{sec: Numerical Examples}.
In this way,
the MAPO model helps us to understand the complicated nature of CPS.

Unlike conventional phase descriptions,
the proposed phase description does not reduce
the number of degrees of freedom of the model,
because we need to iterate the $\left(m - 1\right)$-dimensional map $\eqref{eq: reduced map dynamics}$
to generate the natural measure $\mu_i$.
(Recall that $m$ denotes the number of degrees of freedom of the full model.)
Nevertheless, the MAPO model has
the advantage that in comparison to the full model, it provides a concise description of the rhythm dynamics,
and for this reason, it clearly elucidates how the rhythmic properties of the chaotic oscillator are determined,
as we now describe.
First, the chaotic map $\eqref{eq: reduced map dynamics}$ describing the non-phase dynamics
generates an orbit on the surface of section,
which provides the natural measure $\mu_i$ under the perturbation.
Next, this natural measure determines
the properties of the phase oscillator $\eqref{eq: underlying phase oscillator}$ via Eq.~$\eqref{eq: adjuster}$.
Finally, the averaged equation $\eqref{eq: flow dynamics}$ derived from this phase oscillator
determines the long-term rhythmic properties of the chaotic oscillator.
For most chaotic oscillators,
such a clear interpretation cannot be extracted from the full models.
In addition, the MAPO model has the practical advantage that with it,
we can obtain the Arnold tongue with a very short computation time,
because even the most time-consuming step consists merely of
the iteration of the map $\eqref{eq: reduced map dynamics}$ to generate the natural measure $\mu_i$.
In this regard, it is important to note that
we do not need to recalculate
$T(i)$, $R_\text{r}\bigl(T(i);\,i\bigr)$,
$\Phi\bigl(T(i);\,i\bigr)$, $\Upsilon(t;\,i)$,
and $\zeta(\phi;\,i)$
every time the values of the parameters of the periodic driving are changed,
because these functions are independent of
the amplitude, frequency, and functional form of the periodic driving.

Although in this paper
we have restricted the class of perturbations to external driving functions,
preliminary results suggest that
the rhythm dynamics of mutually interacting chaotic oscillators
also can be described with the MAPO model.
Further analysis of this point will be conducted in the near future.

\begin{acknowledgments}
This work was supported
by MEXT KAKENHI Grant Numbers 15H05877, 16H01617, 18H04948, and 25120011,
and by JSPS KAKENHI Grant Numbers
19H04183, 20H04144, 20K20520, and 20K21810.
\end{acknowledgments}


%

\end{document}